\documentclass[epj,final]{svjour}
\usepackage[T1]{fontenc}
\usepackage[latin1]{inputenc}
\usepackage{graphicx}


\begin{document}

\title{From chaos of lines to  Lagrangian structures in flux conservative fields}

\author{Xavier Leoncini\inst{1,2} \and Olivier Agullo\inst{2} \and Magali Muraglia\inst{2} \and Cristel Chandre\inst{1}}

\institute { Centre de Physique Théorique\thanks{Unit\'{e} Mixte de Recherche (UMR 6207) du CNRS, et des universit\'{e}s
Aix-Marseille I, Aix-Marseille II et du Sud Toulon-Var. Laboratoire
affili\'{e} \`{a} la FRUMAM (FR 2291).%
Laboratoire de Recherche Conventionné du CEA (DSM-06-35).}, Université de Provence-CNRS
Luminy, Case 907 F-13288 Marseilles cedex 9, France \email{Xavier.Leoncini@cpt.univ-mrs.fr}  \and PIIM, Université de Provence-CNRS, Centre Universitaire de
Saint Jérôme, F-13397 Marseilles, France \email{Olivier.Agullo@up.univ-mrs.fr}}

\abstract{ A numerical method is proposed in order to track field
lines of three-dimensional divergence free fields. Field
lines are computed by  a locally valid Hamiltonian mapping,
which is computed using a symplectic scheme. The method is theoretically
valid everywhere but at points where the field is null or infinite.
For any  three dimensional flux conservative field for which problematic points are sufficiently
sparse, a systematic procedure is proposed and implemented. Construction of field
lines is achieved by means of tracers and the introduction of various Hamiltonians
 adapted to  the ``geometrical state'' each line or tracer is. The states
are artificially defined by an a priori given frame of reference
and Cartesian coordinates, and refer to a Hamiltonian which is 
locally valid at the time step to be computed. This procedure ensures
the preservation of the volume (flux condition) during the iteration.
This method is first tested with an ABC-type flow. Its benefits when
compared to typical Runge-Kutta scheme are demonstrated. Potential
use of the method to exhibit {}``coherent'' Lagrangian structures
in a chaotic setting is shown. An illustration to the computation
of magnetic field lines resulting from a three-dimensional MHD simulation
is also provided.}

\maketitle

\section{Introduction}

With the constant increase of computing power, one is now able to
simulate more and more precisely complex systems, among which three
dimensional fluid flows. In order to tackle these simulations one
is often bound to use an Eulerian perspective on flows or fields.
However, if one is interested in transport properties, a Lagrangian
perspective is often best suited \cite{Chernikov90}. For instance
the phenomenon of chaotic advection translates the fact that fluid trajectories
are chaotic despite the laminar structure of the flow \cite{Aref84}.
In two dimensional flows chaotic advection has been studied extensively,
as it offers the possibility to drastically increase the mixing properties
of a given time-dependent flow. If the flow is stationnary  trajectories   of passive tracers resume to field lines,
and no chaos or mixing occurs in two dimensions. However when considering three-dimensional fields, 
field lines are generically chaotic for a stationary flow. Understanding the field line chaotic structures
 in these stationary flows can be therefore  considered as a first step towards understanding transport in these flows. Note also that anyhow, a passive tracer trajectory is locally tangent to the field line for the considered time, hence computing field lines correctly is as well a first numerical step in order to compute trajectories.
 This  chaos of field lines  has been investigated
for a long time in plasma physics, especially when concerned with
the conception of magnetically confining devices such as tokamaks
\cite{White01}. Indeed, the conception of such devices originates from the fact that charged particles are locally trapped by and along magnetic field lines. The chaos of field lines is therefore a crucial ingredient for the understanding and  characterisation of  plasma transport properties.
Note although that in plasma devices, such that a tokamak, numerical
studies of three dimensional magnetic field lines  is simplified
thanks to the toroidal geometry and a strong anisotropic magnetic
field.

In this paper we propose a volume (flux) preserving numerical approach to
the computation of field lines of three dimensional divergence free
fields, while making no assumption on the anisotropy and geometry
of the field. A first attempt in this direction has been proposed in
\cite{Qin93}, however such schemes are first order ones and 
are not generic to flux conservative fields.
Field lines of conservative flux fields as we already mentionned,  known to
be generically chaotic in three dimensions (see for instance \cite{Govorukhin99}
for some nice illustrations and references). This property is often
illustrated by  means of a special local transformation: the equation
of field lines can be written in a Hamiltonian form. These Hamiltonians
belong to the class of $1$-$\frac{1}{2}$ degrees of freedom systems,
which are known to exhibit generically chaotic behaviour. More specifically the field
lines exhibit Hamiltonian chaos, implying some volume preserving constraints as well as ``time-reversibility''.
These features make Hamiltonian chaos quite peculiar, no source nor sinks are possibles, neither are attractors, as such the dynamics does not have special asymptotics displaying for instance a fractal attractor. 

Our strategy to compute field lines numerically while keeping these specific chaotic features
is to use this Hamiltonian mapping and couple it to a symplectic scheme. The symplectic scheme ensures the
preservation of the flux of the original field, by preserving the
volume in phase space \cite{Kang95}.  We insist, that this volume preservation is
essential in Hamiltonian dynamics as it prevents the system from having
any attractor as well as source or sinks in phase space, a feature
which is essential when considering transport properties for large
times and computing the kinetics limit. In the same spirit, we may
expect that paying attention to  volume preservation while computing
field lines becomes quite relevant when the field itself is "properly"
computed. We might think therefore to couple this numeric scheme for
field lines to an exactly conservative integrator for the Euler flow,
such as the one discussed in \cite{Shadwick99}. Finally we would like to mention
that the full three dimensional incompressible and ideal hydrodynamics can be also reduced
to a Hamiltonian formalism \cite{Antoniou04}, using for instance Clebsch variables, but we are then dealing 
with an infinite dimensional phase space, the numerical algorithms are then not easily implemented although
using the so called ``vortex-line representation''  seems promising \cite{Kuznetsov06}. Our approach is much more modest as we do not deal with the computation of the field itself, but also allows us to consider any divergence free field, such as a magnetic field for instance.

Another potential interest in computing field lines for a flux free
field is the possibility to compute Poincar\'e sections of the field
lines. We are then able to represent global information regarding
the three dimensional field on a plane or a number of planes, offering
de facto another perspective on the flow and a tool from which physical
phenomena may have a simpler or different explanation. In fact, this
point of view has been recently used quite successfully in the context
of identification of three-dimensional reconnective structures of
a plasma in a peculiar geometry \cite{borgogno05}, which were ill-defined
in a Eulerian context. With such an approach, we should also be able
to discriminate the physical importance of islands of regular motion
within a stochastic sea in the Poincar\'e sections: For instance, by
identifying the border between chaotic and non-chaotic zones, we might
identify a localised three dimensional structure which does not necessarily
have an Eulerian counterpart. If the Eulerian counterpart exists,
we might have a tool to define clearly its shape and we might test,
as another step, its dynamic robustness. 

The paper is organised as follows. In Section~\ref{sec:Basic-Equations}
a brief summary of field line equations and the Hamiltonian mapping
is given. In Section~\ref{sec:Application-to-numerical}, we present
how the Hamiltonian formalism is applied to numerical data. Then in
Section~\ref{sec:Applications}, we apply the method to two different
examples: First an ABC like flow is considered. The field lines are
computed with a symplectic scheme and the Hamiltonian formalism and
are compared with a Runge-Kutta integration, short-comings of the
latter are clearly exhibited. A few different choices of parameters
are then chosen for the ABC flow, and some three dimensional structures
are extracted. Finally as a proof of feasibility beyond toy models,
the method is applied to a magnetic field computed directly from an
MHD simulation and Poincar\'e sections are shown.

\section{Basic Equations\label{sec:Basic-Equations}}

\subsection{Equation of field lines}

Let us briefly recall the definition of field lines, for this purpose
let $\mathbf{v}$ be a three-dimensional vector field. Field lines
of $\mathbf{v}$ are curves which are tangent to the field at any
point. This definition may be problematic when the field is zero valued
at one point, however we may expect for a reasonable smooth field
that these points if they exist may live on a subset of zero measure
(their union have a zero volume). In the following we will consider
that the computed field lines do not cross such a point. In a mathematical
sense we may consider the definitions of field lines as:\begin{equation}
\mathbf{v}\wedge d\mathbf{M}=\mathbf{0}\:,\label{eq:Field_line}\end{equation}
where $d\mathbf{M}$ stands for a small displacement along the field
line around a point $M$. It is usually easier to consider a given
reference frame and use coordinates, we then rewrite Eq.(\ref{eq:Field_line})
as

\begin{equation}
\frac{dx}{v_{x}}=\frac{dy}{v_{y}}=\frac{dz}{v_{z}}=\frac{ds}{v}\label{eq:Field_line_parametric}\end{equation}
where $dx$, $dy$, $dz$ are the coordinates of the displacement
in a given frame, and $v_{x}$, $v_{y}$, $v_{z}$ are the coordinates
of $\mathbf{v}(M)$ and $v$ is its norm; $ds$ stands for the norm
of $d\mathbf{M}$ (we assumed an orientation of the line is given
). Let us consider now a smooth field, we then choose our coordinate
systems such that in a given region $v_{z}\ne0$. Equations (\ref{eq:Field_line_parametric})
are directly reduced into a system of two ordinary differential equations\begin{equation}
\left\{ \begin{array}{ccc}
dx/dz & = & v_{x}/v_{z}\\
dy/dz & = & v_{y}/v_{z}\end{array}\right.\:,\label{eq:Field_lines}\end{equation}
where $z$ acts as a time variable. Note that this reduction is only
locally valid as it may induce some fictitious singularities if the
line ends up in the neighbourhood of points where $v_{z}=0$. This
point will be discussed again later (see sec.\ref{sec:dealing-with}). Let us now focus on the Hamiltonian
formalism.

\subsection{Hamiltonian formalism for divergence free fields\label{sub:Hamiltonian-formalism}}

We are interested in finding a transformation which will allow one
to describe the evolution of tracers along a field line in a Hamiltonian
formalism. So let $\mathbf{v}$ be a three dimensional divergence
free field, such as for instance the velocity field of an incompressible
flow or a magnetic field and let us reformulate Eq.(\ref{eq:Field_lines})
into a Hamiltonian formalism in this frame. The divergence free condition

\begin{equation}
\mathbf{\nabla\cdot v}=0\:,\label{eq:div_eq_zero}\end{equation}
implies the existence of a potential vector $\mathbf{\xi}$ such that\begin{equation}
{\bf \nabla}\wedge{\bf \xi}=\mathbf{v}\:,\label{eq:xi_def}\end{equation}
\textbf{$\mathbf{\xi}$} being defined up to a given arbitrary gradient
(gauge condition).

Let us consider the vector potential $\mathbf{\xi}$ and relabel its
coordinates $\xi_{x}=-p$, $\xi_{z}=H(x,y,z)$. Now using the gauge liberty
on the vector potential $\mathbf{\xi}$, we set $\xi_{y}=0$. Given
Eq.(\ref{eq:xi_def}) the field $\mathbf{v}$ is rewritten as follows
\begin{equation}
\mathbf{v}=\left(\frac{\partial H}{\partial y},-\frac{\partial H}{\partial x}-\frac{\partial p}{\partial z},\frac{\partial p}{\partial y}\right)\:.\label{eq:speed_expression}\end{equation}
Notice taht the equation of motion for the Hamiltonian $H(x,y,z)$ are non-canonical. In what follows, we change the coordinate system such that they become canonical.
 Given the coordinates $\left(x,y,z\right)$, we define $q=x$ and
$\tau=z$ and switch to the new system $\left(q,p,\tau\right)$. We
then consider the function $\widetilde{H}(p,q,\tau)=H(x,y,z)$. Before
moving on, we recall that the change of coordinates implies\begin{equation}
\left\{ \begin{array}{ccc}
\partial_{x}f & = & \partial_{q}\tilde{f}+\partial_{x} p\partial_{p}\tilde{f}\\
\partial_{y}f & = & v_{z}\partial_{p}\tilde{f}\end{array}\right.\:,\label{eq:deriv_transform}\end{equation}
where $\partial_{a}f=\partial f/\partial a$. We then rewrite the
equations of the field (\ref{eq:Field_lines}) in the new coordinate
system and readily obtain from Eqs.(\ref{eq:speed_expression})\begin{equation}
\left\{ \begin{array}{ccc}
\dot{q} & = & \frac{\partial H}{\partial p}\\
\dot{p} & = & -\frac{\partial H}{\partial q}\end{array}\right.\:,\label{eq:Hamilton_eq}\end{equation}
where the dot refers to a derivative with respect to $\tau$ and the
tilde was removed from $H$. The equations are of canonical Hamiltonian form
with the $(p,q)$ acting as canonical conjugates, and $\tau$ acting
as a time variable. Hamiltonian $H$ is the component of the potential
vector associated with the chosen fictitious time (the $z$-direction
in our case).

\subsection{Locality\label{sub:Locality}}

Note that the singularity for $v_{z}=0$ is not apparent in Eq.(\ref{eq:Hamilton_eq}).
However the singularity remains because the fictitious time becomes
singular for $v_{z}=0$, which if not cared about might reverse the
arrow of "time". This transformation is actually only locally valid.
However, first, given our hypothesis of a relatively smooth field,
it is unlikely that we hit a $\mathbf{v=0}$ point. Second, we have
a free choice of our coordinate system, thus we can rotate our frame
as we think is best. With this freedom of transformation, a natural
choice would be to consider a locally well suited coordinate system in order
to avoid singularities in $v_{z}$. Unfortunately, this strategy is
not well suited when computing many field lines numerically at the
same time. Indeed each would need its own local Hamiltonian since
the gauge condition will be point dependent. And, as a result this
would be numerically expensive.

Besides the singularity, injectivity of the transformation $\left(x,y,z\right)\rightarrow\left(q,p,\tau\right)$
must be ensured if one wants to perform time evolution in Hamiltonian
variables. Indeed, the momentum is formally defined by\begin{equation}
p=\int^{y}v_{z}dy\:,\label{eq:formal_p_of_y}\end{equation}
and, without more specifications, unicity in the determination of
$y$ given the triplet $(q,p,\tau)$ is not guaranteed: the history
of $v_{z}$ along the field line is included in $p$ and we may recover
problems dealt with previously, related to $v_{z}=0$. Moreover, it
is also obvious in Eq.(\ref{eq:formal_p_of_y}) that another difficulty
will arise in a region for which $v_{y}=0$. In this region we expect
to have $dy=0$ and thus a stationary $p$ in $y$, making the transformation
likely non-invertible. 

As will be discussed in Sec.~\ref{sec:Application-to-numerical},
we settled for a compromise between implementability as well as numerical
cost and debugging time. Let us now discuss one of the key reasons
for this transformation to be performed.

\subsection{Volume preserving scheme}

We will now use this Hamiltonian structure in order to compute numerically
the field lines of $\mathbf{v}$, because it will allow the use of a
symplectic scheme which ensures the preservation of invariants, and which
given the divergence free condition (\ref{eq:div_eq_zero}), implies the
conservation of volume along field lines.

It is well known that intrinsically to any source-free systems, there is a
volume form of the phase space $\Re^{n}$, say\begin{equation} f=dx_{1}\wedge
dx_{2}\dots\wedge dx_{n}\;, \label{eq:form_symplectic}\end{equation}
such that the dynamic evolution preserves this form. This constraints the
geometric structure of the field lines. To explicit the structure
numerically, it is crucial to use a scheme preserving this property as shown
in section \ref{sec:Application-to-numerical}. 

For $n=2$, source free vector fields are Hamiltonian fields and symplectic
algorithms are area-preserving maps. The situation is different for $n\ge3$:
all the conventional methods, including the well-known Runge-Kutta and Euler
methods, are non volume preserving (see \cite{Kang95,Hairer_book}).
By means of an {}``essentially Hamiltonians decomposition'' of source free
vector fields, a volume preserving difference scheme was proposed in
Ref.~\cite{Kang95}. The algorithm we used - we will restrict ourselves to
the case $n=3$ - does not use the decomposition of Ref.~\cite{Kang95}:
in such a way, computation of the {}``time evolution'' of the tracers along
the field lines requires the explicit introduction of only one Hamiltonian
(and not two), the price being the introduction of a time dependent
preferential space direction (see Sec.~\ref{sub:Locality}).
However, the idea is still to use the underlying \textit{local} Hamiltonian
structures, as explained in Sec.~\ref{sub:Hamiltonian-formalism}, and
integrate the dynamical equations (\ref{eq:Hamilton_eq}), by means of
symplectic schemes. This procedure, indeed, allows the preservation of the
volume form 
\begin{equation}
\label{eqn:pres}
dx\wedge dy\wedge dz=dq\wedge dp\wedge \frac{d\tau}{v_z}.
\end{equation}
The meaning of Eq.~(\ref{eqn:pres}) is the following one~: We know that
$dx\wedge dy\wedge dz$ is preserved as time $t$ evolves. From the
Hamiltonian framework we have that $dq\wedge dp\wedge d\tau$ is preserved as
the fictitious time $\tau$ evolves which is here the coordinate $z$. The
consistency is provided by the way we follow the evolutions, i.e., by the
nonlinear relation $dz/dt=v_z$ between these two evolutions.

\section{Application to numerical data\label{sec:Application-to-numerical}}

\subsection{Fourier representation}

We now present a possible way to implement this formalism when dealing
with a field obtained by numerical simulation. We consider a quite
common case of a field with periodic boundary conditions. If this
is the case, it is a common practice to consider the Fourier space for
the computation as for instance when one uses a pseudo-spectral method.
We therefore consider a case where all the Fourier components ${\bf \hat{\mathbf{v}}_{{\bf k}}}$
of $\mathbf{v}$ are known. Here the vector ${\bf k}$ is used as
a mode index. We shall also for convenience consider that the number
of modes is finite, which is more realistic when dealing with numerical
data. Using equations (\ref{eq:speed_expression}) we then write the
Hamiltonian and momentum in the $(x,y,z)$ coordinates: \begin{eqnarray}
H(x,y,z) & = & \sum_{k_{y}\ne0}\frac{\hat{v}_{\mathbf{k},x}e^{i{\bf k}\cdot{\bf x}}}{ik_{y}}+y\left(\sum_{k_{y}=0}\hat{v}_{\mathbf{k},x}e^{i\mathbf{{\bf k}}\cdot{\bf x}}\right) \\
 &  & -\sum_{k_{y}=0,k_{x}\ne0}\frac{\hat{v}_{\mathbf{k},y}e^{i{\bf k}\cdot{\bf x}}}{ik_{x}}-x\left(\sum_{k_{y}=0,k_{x}=0}\hat{v}_{\mathbf{k},y}e^{i{\bf k}\cdot{\bf x}}\right)\label{eq:H_of_y}\\
p & = & \sum_{k_{y}\ne0}\frac{\hat{v}_{\mathbf{k},z}e^{i{\bf k}\cdot{\bf x}}}{ik_{y}}+y\left(\sum_{k_{y}=0}\hat{v}_{\mathbf{k},z}e^{i{\bf k}\cdot{\bf x}}\right)\:.\label{eq:p_of_y}\end{eqnarray}
Note that the condition $\xi_{y}=0$ is not fully sufficient to determine
$H$ and $p$, and the equations (\ref{eq:H_of_y}) and (\ref{eq:p_of_y})
which are defined up to a function of $x$ and $z$, reflect our choice
of trying to keep $p$ as simple as possible, leaving the {}``complexity''
in $H$.

\subsection{Numerical scheme}

In order to compute field lines, we take advantage of the Hamiltonian
formalism and use a symplectic scheme. In the present case we consider
the fourth and sixth order Gauss-Legendre scheme proposed in \cite{McLachlan92}.
However the transformation from the $(x,y,z)$  to $(p,q,\tau)$
 is not trivial. For instance, having access to the function $\widetilde{H}(p,q,\tau)$
in order to compute the Hamiltonian dynamics seems rather complex.
In order to avoid these difficulties, even though the field lines
are propagated in the $(p,q,\tau)$ space, we decided to compute the
$-\partial_{q}\tilde{H}$ in the $(x,y,z)$ space using the expressions
(\ref{eq:deriv_transform}).

We therefore have to keep a constant correspondence between the two
spaces. This is performed by considering the expression of $p$ given
by Eq.(\ref{eq:p_of_y}), and explains our gauge choice to keep it
as simple as possible. In order to get $(x,y,z)$ given a value of
$(p,q,\tau)$, only $y$ is given implicitly by Eq.(\ref{eq:p_of_y})
since $x=q$ and $z=\tau$. The inversion is done numerically using
a simple Newton scheme, for which the first guess for the root is
given by the Euler method.

\subsection{Dealing with pseudo-singularities\label{sec:dealing-with}}

Up to now we have assumed that the Hamiltonian transformation is globally
valid. By the choice of coordinate system, the direction $z$ acts
as a time variable, and is therefore assumed to be a monotonic function
along the field line. One may therefore wonder what happens when we
hit a fold, which turns the field line in the reversed direction.
If this is the case $z$ can not be monotonic and the Hamiltonian
transformation fails. Our way of dealing with such problems is to
assign a state to each field line, corresponding to which direction
the line is propagated in. Indeed when we reach a point where $v_{z}=0$,
one may expect that $v_{x}$ or $v_{y}$ is not null. Since our coordinate
system is arbitrary, we may conceptually rotate our frame in order
to get $v_{z}\ne0$ in this new coordinate system and propagate our
field with the symplectic scheme. Ideally, it would be best suited to locally
rotate the coordinate system and try to take the arc-length along the line
as a measure of the fictitious time variable. However when computing many field
lines together, this procedure would mean to keep track of all coordinate
systems and rotate them after each step to the local Frenet system
before performing the transformation and the next time step. This
seems numerically expensive. We settled for a
more pragmatic point of view. Field lines are thus assigned a number
$i\in\{1,2,3\}$ corresponding to which coordinate system the Hamiltonian
transformation is applied to. These numbers corresponding to a cyclic
permutation of the coordinates $(x,y,z)$ of a coordinate system given once and
for all.

Furthermore as was previously discussed, problems may also arise when
$v_{y}=0$. If such a seldom situation occurs, we simply permute the
current $x$ and $y$ directions. Formally this corresponds to a different
choice for $p$ and $q$ and a different choice of gauge for $\mathbf{\xi}$.
A new number $j\in\{1,2\}$ is therefore assigned to the field line.

When changing the coordinate system, we have to make sure that we
will continue to {}``advance'' along the line, meaning we have to
be careful and make sure that the new Hamiltonian transformation will
not end up in going backtrack along the already computed path. We therefore
choose an orientation for the lines and impose the condition \begin{equation}
\mathbf{v}\cdot d\mathbf{M}>0\:.\label{eq:field_line_orientation}\end{equation}
 It is obviously important to ensure that this condition is still
satisfied when changing coordinates. Since it is conceptually easier
to deal with a positive time step $d\tau$, and since field lines
are invariant when multiplied by a scalar field, we perform the change
$\mathbf{v}\rightarrow-\mathbf{v}$ when necessary in order to comply
with Eq.(\ref{eq:field_line_orientation}). This implies that another
number $k\in\{1,2\}$ must be assigned to the field line. 

At last, since we try to avoid performing a rotation of our coordinate
system, except for simple permutation of the coordinates, we may hit
points for which the field is along one of the coordinates, leaving
us with two zero-valued coordinates and problems of invertibility.
In order to deal with such situations, we actually have to consider
a second reference frame, not issued by a permutation. We may then
consider for instance a $\pi/4$ rotation around the $z$-direction
of the original coordinate system, and deal with problematic points
in this frame. Given all possibilities while computing the field line,
its state (the numbers $(i,j,k)$) can thus take twelve different
values, and we can be in two different coordinate systems, giving
us a broad range of twenty four Hamiltonians in order to compute a
field line.

\section{Applications\label{sec:Applications}}

In order to test the algorithm and start to explore the possibilities
offered by the analysis of field line chaos, we started with a basic
three dimensional flow, namely the ABC-flow \cite{Arnold65,Childress70}.
We first validate our numerical scheme by considering a choice of
parameters for which no singularity occurs and the Hamiltonian transformation
is global. Further on, we show the advantages of using a symplectic
scheme over a non-symplectic one. We then consider integrable and
chaotic flows for which singularities occur. In the latter case, Lagrangian
structures are exhibited as a result of the analysis of Poincar\'e sections.
Finally as a proof of concept we compute the magnetic field lines,
in the context of the dynamo process, of a magneto-hydrodynamic flow.

\subsection{ABC flow}

In order to test the algorithm we consider the following periodical
flow which may be seen as a particular example of the well know ABC
flow \cite{Arnold65,Childress70}:\begin{eqnarray}
v_{x} & = & \cos y-\varepsilon\sin z\nonumber \\
v_{y} & = & \sin x+\varepsilon\cos z\label{eq:abc_flow}\\
v_{z} & = & \cos x-\sin y+v_{0z}\:,\nonumber \end{eqnarray}
where $\varepsilon$ and $v_{0z}$ are parameters. 

\begin{figure}
\includegraphics[%
  width=8cm,
  keepaspectratio]{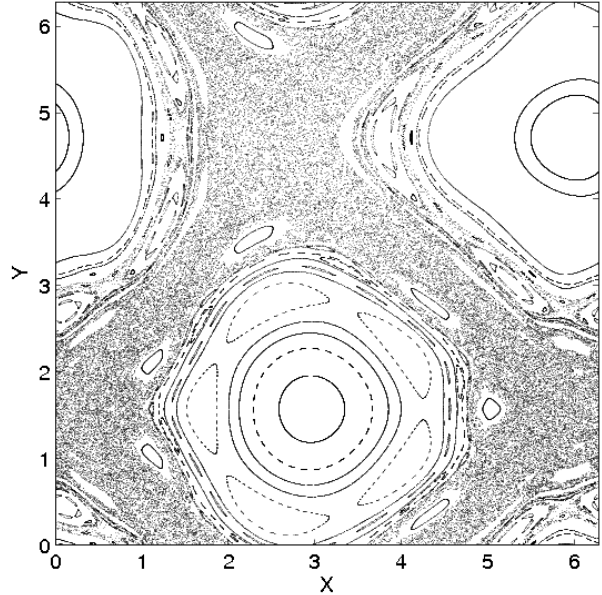}

\caption{Poincar\'e section of the flow (\ref{eq:abc_flow}) with $v_{0z}=4$ and $\varepsilon=0.9$.
The Poincare section is performed on the $z=0\ \:[2\pi]$ plane\label{cap:Poincar=E9-map-of 1}.
The flow is computed using the 4th-order symplectic scheme with a
time step $\delta z=\pi/100$. }
\end{figure}

We started to benchmark the code by comparing its results with the
one obtained using a non symplectic fourth order Runge-Kutta scheme
on Equations (\ref{eq:Field_lines}). Note that in order to have some
consistency, comparison is done with the symplectic fourth-order Gauss-Legendre
integrator \cite{McLachlan92}.  For this first test we took $\varepsilon=0.9$
and $v_{0z}=4$ in order to avoid singularities which occurs at points
where $v_{z}=0$. Since  $v_{z}>0$, $z$ is used as the time variable, and due to the
periodic character of the flow, we can reduce our analysis of the flow on the  Poincar\'e map defined by  the crossing of field lines with the successive $z=0\:[2\pi]$ planes.
The resulting plot is  depicted in Fig.~\ref{cap:Poincar=E9-map-of 1}
using the two methods.
The {}``time step'' used for the run was $\delta z=\pi/100$.
To the naked eye, the same section is obtained for both cases, points
solely differ in the chaotic region. The chaotic behaviour of steady
field lines in three dimensions is clear. This first test is used in order to confirm that the proposed method
computes what it is expected to.

 Now let us consider the advantages of using such a symplectic algorithm.
For this purpose we consider the integrable case $\varepsilon=0$, and we compare the
evolution of a field line starting from the same initial condition (for visual
purposes) but computed using both methods. Since the flow is integrable we expect to see a closed line on the Poincaré map. %

\begin{figure}
\includegraphics[%
 clip, 
  width=8cm,
  keepaspectratio]{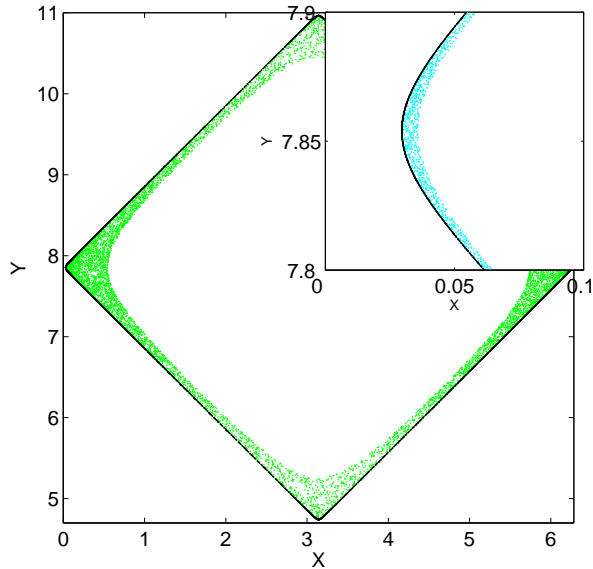}

\caption{Poincar\'e section of the flow (\ref{eq:abc_flow}) with $v_{0z}=3$ and $\varepsilon=0$.
The Poincare section is performed on the $z=0\ \:[2\pi]$ plane for
only one initial condition $(3.142,4.742)\sim(\pi,3\pi/2)$. The black
line corresponds to the expected quasi-periodic trajectory obtained with the
fourth-order symplectic scheme. The cloud of green dots corresponds to the
trajectory obtained by the fourth-order Runge-Kutta scheme, showing  some diffusion towards the centre of the island. Integration
is performed up to $z=\pi\:10^{6}$, with a time step $\delta z=\pi/10$.
In the insert the same simulation is performed for the Runge-Kutta
scheme but with a smaller time step $\delta z=\pi/20$ diffusion is still present, the dark line being
a replot of the results obtained with the symplectic scheme and the
larger time step $\delta z=\pi/10$.\label{cap:Rk4-versus-LagHam}}
\end{figure}

First we consider both cases with an identical relatively large time
step $\delta z=\pi/10$. The results are shown in Fig.~\ref{cap:Rk4-versus-LagHam}
and clearly demonstrate that  the non symplectic method has some shortcomings as the expected closed line becomes a cloud of points,
and inward diffusion is observed. Now in order to be "fair", since the non symplectic scheme is almost twice faster on an
identical computer than the {}``conserving'' one, we took a smaller
time step $\delta z=\pi/20$ for the Runge-Kutta method in order to
match the integration times (CPU times) between the methods. Still,
as seen in the insert in Fig.~\ref{cap:Rk4-versus-LagHam} some inward
diffusion is observed with the Runge-Kutta scheme. These non-conservative features
may induce spurious effects, for instance when considering transport properties in the kinetic limit,
while transport may be not Gaussian (see for instance \cite{leoncini05,leoncini01}) one may end observing a diffusive behaviour with a diffusion coefficient governed by the algorithm, as this diffusive behaviour will end up erasing in a finite time any potential memory effect generated by the presence of regular trajectories in a mixed phase space. In this sense using a conservative algorithm despite its extra numerical overhead, we may avoid false interpretations which may arise with Runge-Kutta as for instance possible transient anomalous behaviour.

In the rest of the paper we use the sixth-order Gauss-Legendre scheme.

\subsection{Flow with pseudo singularities}

\begin{figure}
\includegraphics[%
  width=4cm,
  keepaspectratio]{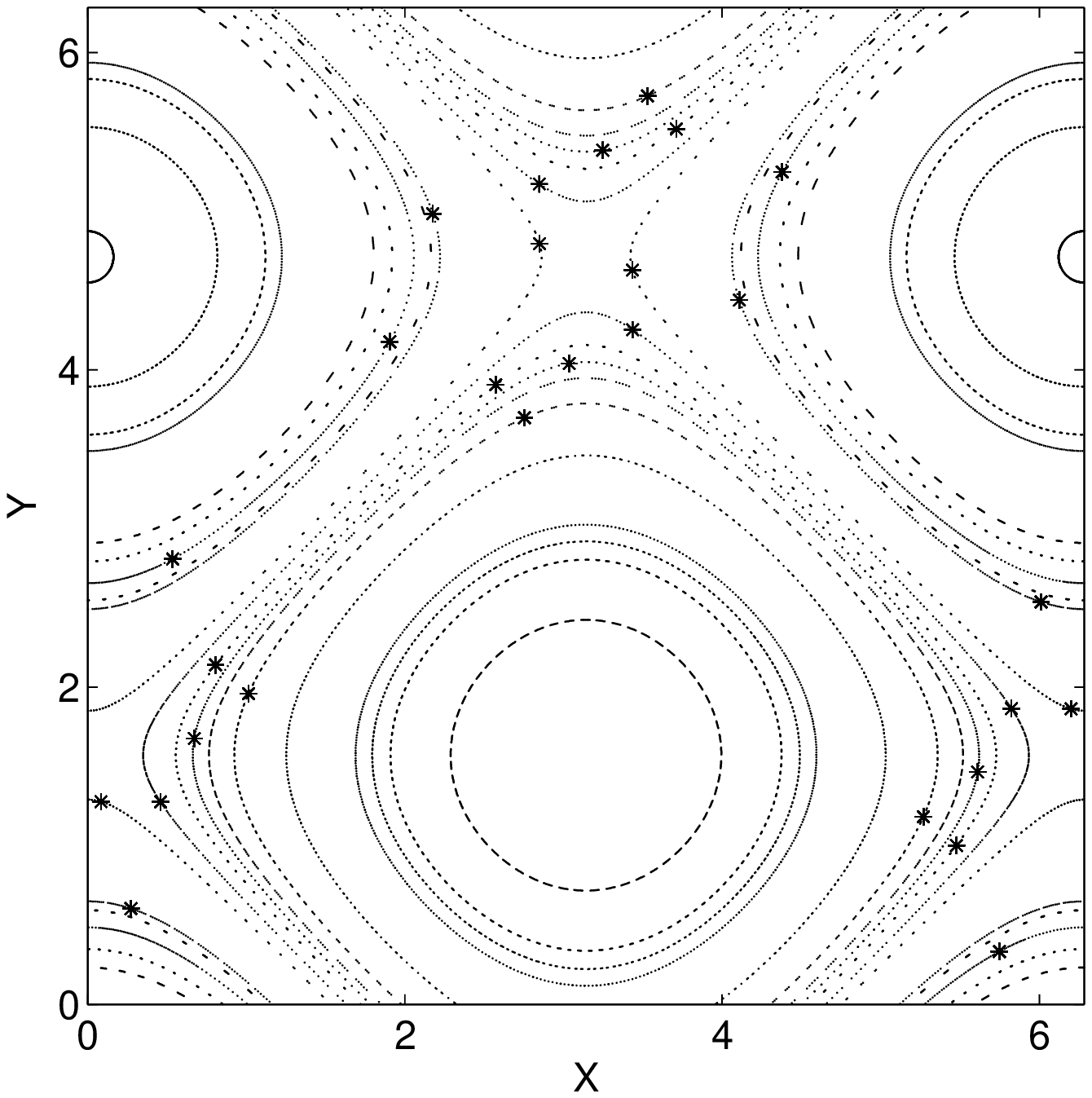}\includegraphics[%
  width=4cm]{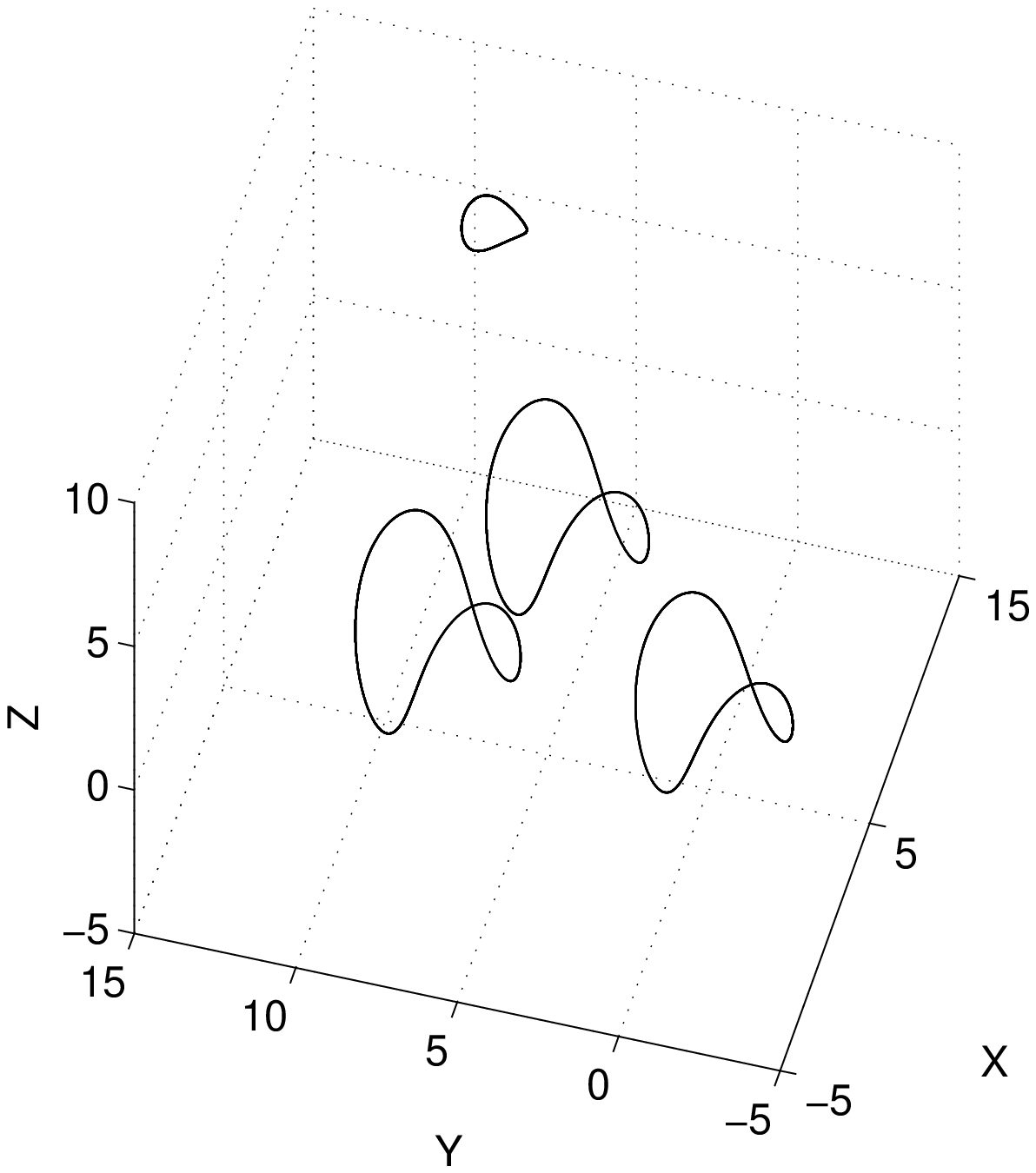}

\caption{Left: Poincar\'e section of the flow (\ref{eq:abc_flow}) for the integrable case $\varepsilon=0$:
note that the same section is obtained as we take different values
of $v_{0z}$ ranging from $3$ to $0.01$ ($v_{0z}\rightarrow0$).
The {*} signs, correspond to points obtained for $v_{0z}=0$, the
field lines appear to be periodic for this case. The Poincar\'e section
is performed on the $z=0\ \:[2\pi]$ plane with $\delta\tau=\pi/50$.
Right: Four field lines computed for the integrable case $\varepsilon=0$
and \label{cap:Singular_integrable}$v_{0z}=0$. Field lines are indeed
periodic and form closed lines.}
\end{figure}

We now test the code on a flow for which the problems of singularities
and non invertibility occur. Namely we consider a series of flows
for which $v_{0z}\rightarrow0$. We first consider the integrable
case ($\varepsilon=0$). The results are shown in Fig.~\ref{cap:Singular_integrable}.
It is easy to see from Eqs.(\ref{eq:abc_flow}) that the flow becomes
somewhat two-dimensional, as the $z$ dependence of the flow completely
vanishes. For this {}``two-dimensional'' steady flow $\cos x+\sin y=H_{2D}$
is a constant along the field line and corresponds to the stream function
which acts as a Hamiltonian. This property gives rise to the closed
lines observed in the Poincar\'e section depicted in Fig.~\ref{cap:Singular_integrable}.
Thus for any value $v_{0z}>0$ we expect to have identical sections, which is the case as we take different values
of $v_{0z}$ ranging from $3$ to $0.01$ ($v_{0z}\rightarrow0$) observed in Fig.~\ref{cap:Singular_integrable}.
An additional feature occurs when $v_{0z}=0$, all field lines appear
to become periodic and localised (forming a closed line). In this
regard, we like to pinpoint a specific fact about our choice for
the equations of the ABC flow given by Eqs.(\ref{eq:abc_flow}). In
fact it is usually a common procedure to take $v_{z}=H_{2D}(x,y)$ (see
for instance \cite{Zaslav_book_91,Agullo-zas97}). In the integrable
situation this procedure leads to a constant $v_{z}$, and helical
field lines. In these regards, the choice of the minus sign in the
third equation of Eqs.(\ref{eq:abc_flow}) drastically changes the
behaviour of field lines. From the perspective of testing the numerical
scheme, this choice also proved to be more challenging than usually \cite{Zaslav_book_91}, since the field lines "bite their tails".%

\begin{figure}
\includegraphics[%
  width=8cm,
  keepaspectratio]{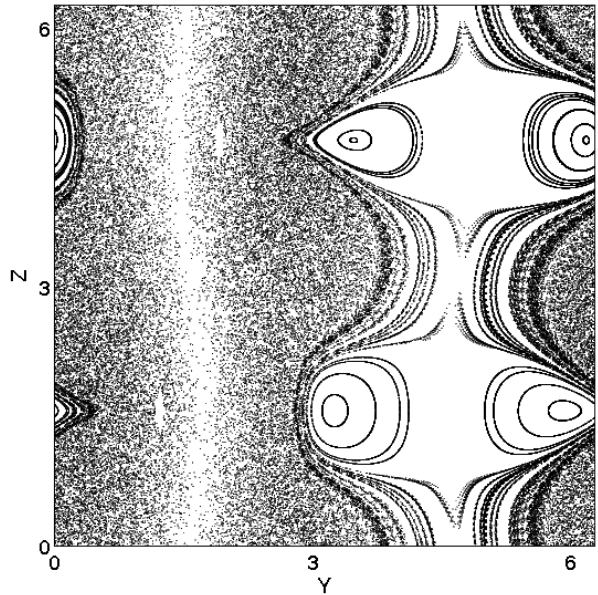}\\

\caption{Poincar\'e section of the flow (\ref{eq:abc_flow}) for a chaotic case $\varepsilon=0.15$,
for a flow with pseudo-singularities. One can infer a chaotic region
and a regular one. One also notices a region which is not often crossed
in the chaotic sea (lighter zone). The Poincar\'e section is performed
on the $x=0\ \:[2\pi]$ plane. The plot is made with $100$ field
lines which are iterated $2\:10^{5}$ times with a time step of $\delta z=\pi/100$.
\label{cap:Poincar=E9-section-of-chaos-o15}}
\end{figure}

Since the numerics are giving us what we expect, we may now move to a non-integrable situation.
For this purpose we consider $\varepsilon=0.15$ with
a $v_{0z}=0$: the integrable case with closed field lines is perturbed.
A Poincar\'e section is shown in Fig.~\ref{cap:Poincar=E9-section-of-chaos-o15}.
One can observe regions with regular quasi-periodic field lines, and
regions of chaos. In the regular region, one can notice some delocalised
field lines, as well as apparently localised ones (at least in the
$z$-direction). In the chaotic sea, one notices a region with a vanishing
density of points, in other words a region which is not often crossed.
Since the depicted section is computed in the $x=0\ \:[2\pi]$ plane,
one can expect that regions in these planes for which $v_{x}=0$ are
not crossed. It implies that regions around the curves given by $\cos y=\varepsilon\sin z$
are not often crossed. These curves correspond to the light region
in Fig.~\ref{cap:Poincar=E9-section-of-chaos-o15} for one, and for
the {}``centre'' of the regular region in Fig.~\ref{cap:Poincar=E9-section-of-chaos-o15}
for the other one. Before concluding on this particular example, we
want to mention the existence of two islands in the chaotic region
nearby the low-density region that the method is able to detect.

\subsection{Lagrangian structures}

In fact the Poincar\'e section shown in Fig.~\ref{cap:Poincar=E9-section-of-chaos-o15}
is quite generic for systems with one and a half degrees of freedom,
with regions of regular motion and a chaotic sea. In this section
we describe the shape of the field lines corresponding to the regular
regions. Indeed contrarily to a true phase space section from a given
Hamiltonian, the section depicted in Fig.~\ref{cap:Poincar=E9-section-of-chaos-o15}
corresponds to field lines obtained typically by a {}``product''
of local Hamiltonians. The {}``time'' dependence is therefore not
always monotonic. We thus may expect more complexity than the typical
braid of helical structures which is expected in the phase-space extended
with a time direction. Besides this possible increase of complexity,
the regular structures may have also a crucial physical impact. For
instance, if one consider the field $\mathbf{v}$ as being the velocity
field of an incompressible fluid, the transport properties of passive
scalars will be along the field lines. Thus the properties of these
regular regions may greatly influence global properties of transport
or mixing in this flow.%

\begin{figure}
\includegraphics[%
  width=8cm,
  keepaspectratio]{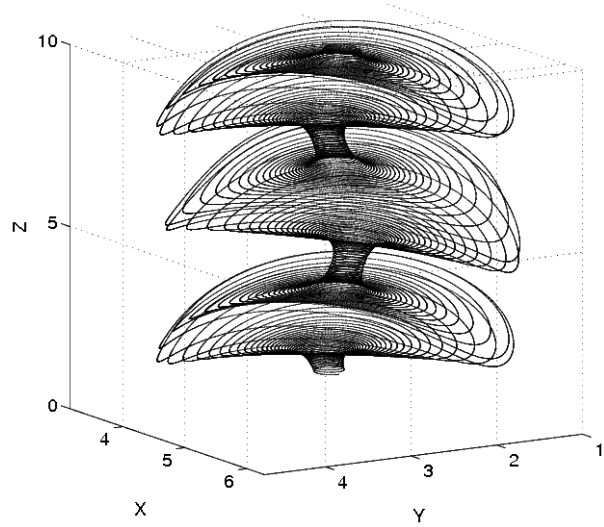}

\includegraphics[%
  width=8cm]{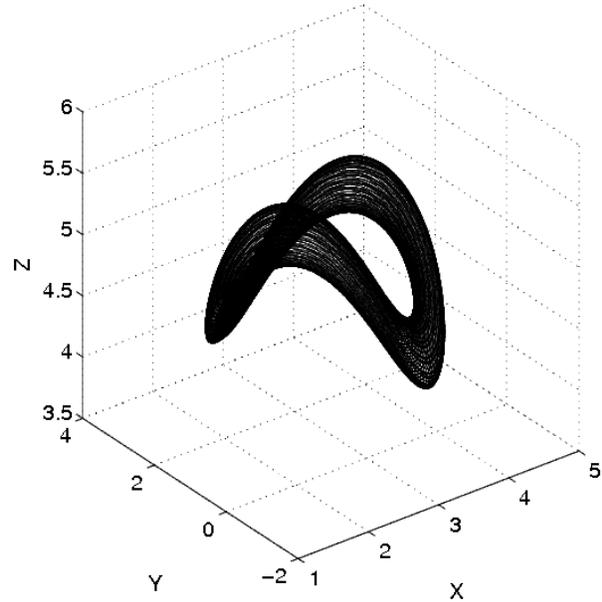}

\caption{Upper panel: regular drifting field line. Lower panel: Localised
field line. \label{cap:Structures Epsilon 0.15}}
\end{figure}

In order to localise the coherent Lagrangian structures associated with regular field lines,
we just consider initial conditions within a regular zone in the Poincar\'e section, and computed
the resulting field lines. To our knowledge this type of methodology has not been applied to generic divergence free fields,
in particular to the considered ABC flow. Results are shown in Fig.~\ref{cap:Structures Epsilon 0.15}.
Typically one notices two types of behaviours, localised field lines,
which are drawing a finite surface, and unlocalised ones. Note the
mushroom-like shape of the unlocalised field line, which if considered
as the trajectory of a passive tracer, may appear as successive sheets
in an experiment. Among the unlocalised regular field lines, one can
also find a simple helical like form, when one considers an initial
condition within the two small islands located not far from the no
crossing region. To conclude on these structures, it is worthwhile
mentioning that even by looking at the sections trough different planes
($y=0\ \:[2\pi]$ or $z=0\ \:[2\pi]$) the same structures are obtained.
They just have different sections with these planes. One may also
infer the Poincar\'e plot depicted in Fig.~\ref{cap:Poincar=E9-section-of-chaos-o15}
that the localised structures (Fig.~\ref{cap:Structures Epsilon 0.15}
lower panel) appear as sandwiched between the unlocalised mushroom
ones (Fig.~\ref{cap:Structures Epsilon 0.15} upper panel).
We may also note that the observed "coherent structures" do not have necessarily a simple tubular shape,
hence we may speculate that using this technique to uncover spatially extended coherent structures in more complex flows
such as MHD, may yields some unexpected shapes with maybe important physical consequences and selection mechanisms.

\subsection{Magnetic Field from MHD}

We shall now consider a full feature numerical study and move on to
the magnetic field obtained through magneto-hydrodynamics (MHD) simulations.
The origin of cosmic magnetic fields can be investigated in the frame
of MHD. The problem is to extract the underlying mechanisms of the
dynamo process, i.e. of the self-excitation of a magnetic field by
plasma or fluid motions. To describe the generation of large-scale
magnetic fields from small-scale turbulence many mechanisms are involved
and, in the following, we will focus on the so-called $\alpha^{2}$-dynamo (see for instance \cite{Branderburg2001}).

\begin{figure}
\includegraphics[%
  width=8cm]{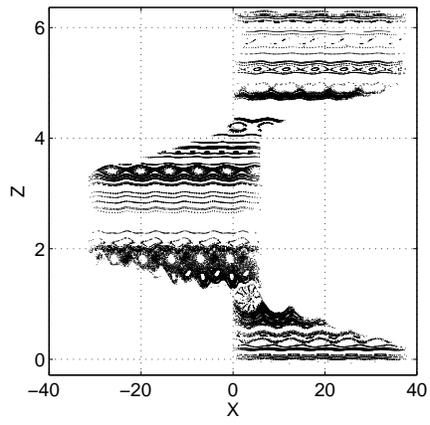}

\caption{Poincar\'e section of magnetic field lines. \label{cap:Poincare_dynamo}}
\end{figure}

The equations of incompressible MHD, written in the usual units, read\begin{eqnarray}
\frac{D}{Dt}{\bf u} & = & -{\bf \nabla p}+{\bf b\cdot\nabla b}+\nu{\bf \nabla^{2}u}+{\bf f}\nonumber \\
\frac{D}{Dt}{\bf b} & = & {\bf b\cdot\nabla u}+\eta{\bf \nabla^{2}b}\label{eq:MHD equations}\\
{\bf \nabla\cdot b} & = & {\bf \nabla\cdot u}=0\:,\nonumber \end{eqnarray}
where by definition, $\eta$ is the magnetic diffusivity, $\nu$ the
kinematic viscosity and the scalar $p$ is the sum of the hydrostatic
and the magnetic pressures. We choose $\nu=\eta=10^{-3}$. Here, both
fluid velocity ${\bf u}$ and magnetic field ${\bf b}$ are expressed
in Alfvèn speed units. The flow is forced by a divergence free field
at scales much smaller than the width $L=2\pi$ of the cubic and periodic
box. $\alpha^{2}$-dynamo is
enhanced by imposing the forcing to be maximally helical: its relative
helicity is one at any time \cite{agullo2003b,Branderburg2001}. Initial
conditions are ${\bf u}=0$ and ${\bf b\sim0}$. The flow is driven
by the forcing and it turns out that the Reynolds number based on
the Taylor microscale is around a few units: this is a moderate turbulent
regime. In fact, a noteworthy effect of the Lorentz force is to decrease by roughly a factor ten the Reynolds value. In Fig. \ref{cap:Poincare_dynamo} are drawn time evolutions of the total kinetic and magnetic energies of the plasma. 
Clearly, owing to the kinetic forcing, kinetic energy is initially rapidly accumulated. The growth of the magnetic energy  initiates then a decrease of the kinetic energy and a quasi-saturated state is reached at time $T=300$. The eddy turnover time is typically of a few time units. Transfer of helicity from the forcing to the flow and from the flow to the magnetic field, but also Alfvén effects and an inverse cascade process (after time $T\sim 15$ here) lead to the generation of a
robust anisotropic large scale magnetic field, characterised by a
pile up of magnetic energy at the largest scales \cite{Biskamp_book,Branderburg2001}.
Once the large scale magnetic field reaches a stationary state, a
rough snapshot of the Eulerian field can be summarised by the approximation
\begin{equation}
{\bf b}\sim\cos z{\bf e_{x}}+\sin z{\bf e_{y}}+\epsilon\sum_{|{\bf k}|\sim k_{F}}{\bf b_{{\bf k}}}\exp(i{\bf k\cdot x})\;,\label{eq:B large scales}
\end{equation}
where $\epsilon$ is a small parameter to indicate that energetically
dominant modes are at the largest scales $k=2\pi/L=1$ and $k_{F}$
is the forcing scale. In Fig. \ref{cap:Poincare_dynamo}, we draw
a Poincar\'e section of the three-dimensional magnetic field lines resulting
from the magnetic field obtained at time $T=900$. This is 
the section of $100$ magnetic field lines. Initial positions of
the tracers to determine the fields lines are randomly distributed
in the periodic box. The computation is done by including in the
field an energetically consequent fraction of the modes. Inclusion
of all the modes would have been too computer time consuming and,
since in this paper we put our effort on showing a proof of concept
and we do not focus on the details of the Lagrangian structures. For
instance, dissipative Eulerian structures are not included in the
structure of the field. The existence of chaotic Lagrangian structures
is however clearly observed in Fig~\ref{cap:Poincare_dynamo}. It
is also observed a``drift of the structures'' according to their
height $z$. In fact, this drift trivially originates from the Eulerian
field as can be seen by inspection of the first term on the r.h.s
of Eq. (\ref{eq:B large scales}). We like to insist
on the fact that Fig.~\ref{cap:Poincare_dynamo} clearly suggests
the existence of a non trivial underlying Lagrangian and chaotic organisation
in duality with the well established coherent Eulerian structures
suggested by the approximation (\ref{eq:B large scales}).
These remarks suggest that the  study of the Lagrangian structures in this type of flow offers promising prospects. Indeed,
chaos of field lines of a flow, neither necessarily turbulent nor unstationnary, is an ingredient known to clearly enhance a dynamo effect; an clear example being the G.O. Roberts dynamo where an ABC flow is used\cite{Gilbert_book2003}. 
The large scale term of the saturated magnetic field in the simulations is, in fact, also of ABC type. However the role of the turbulence in the generation of the former, but also how the inverse cascade generates helical structures in a turbulent flow are not clear. Thinking those processes in terms of Lagrangian structures in parallel with a classification and measure of their helicities might allow a new understanding of the interplay of turbulent, helical and chaotic phenomena in 
dynamo processes. 
Finally, we should
emphasise that the presence of chaos does not necessarily rely onto
any complexity in the small scale perturbative term of Eq. (\ref{eq:B large scales}).
Indeed, it probably essentially results in the presence of the large
scale Beltrami structure, ${\bf b}_{\epsilon=0}=\kappa\nabla\wedge{\bf b}_{\epsilon=0}$
with $\kappa=-1=$Cte, as a consequence of the Arnold theorem\cite{Arnold65}. We may also emphasise on the fact that the existence of chaotic field lines is in turn a signature of the three dimensional nature of the field, and considering the anti-dynamo theorems in two-dimensional flows we may expect that the mechanisms underlying the dynamo effects are generically correlated with the existence of chaotic field lines 

\section{Conclusion\label{sec:Conclusion}}

The main purpose of this work is to propose a way to compute field
lines of divergence free fields, while keeping the conservative flux
condition valid. Field lines are computed by performing a locally
valid Hamiltonian transformation, which in turn is numerically computed
using a symplectic scheme. The method is theoretically valid anywhere
but at points where the field is null or infinite. For fields for
which such points are sufficiently sparse that we consider they numerically
do not exist, a systematic procedure has been proposed. Field lines
are propagated using up to twenty four different Hamiltonians, depending
in which {}``state'' each line is. The states being artificially
defined by an a priori given frame of reference and Cartesian coordinates,
and refers to which Hamiltonian is currently locally valid for the
time step to be computed. The potential application to three dimensional
fields with periodic boundary conditions is discussed and explicit
usable expressions for the Hamiltonian and momentum are given. We
have also provided a detailed possible actual potential implementation
into a numerical code of the ideas, and presented numerical results
obtained for specific systems. In this procedure, we have shown some
advantages of using this Hamiltonian-symplectic formalism over a regular
Runge-Kutta scheme for the field lines.

Moreover, while testing the numerical scheme, field lines of ABC type
flows have been investigated. For a considered integrable flow with
no mean flow, we have shown that field lines can be all closed. Also
a Poincar\'e section of a non-integrable flow, for which the field
lines can not be described by a global Hamiltonian has been computed
and a mixed phase space with a chaotic sea and regular regions is
exhibited. From this analysis Lagrangian coherent structures are extracted,
and either localised or extended structures have been found. Finally
in order to show an application of the numerical tool in order to
tackle physical problems, the field lines of the magnetic field resulting
from an magneto-hydrodynamic simulations in a dynamo process have
been computed. 

\begin{acknowledgement}
We would like to thank G. M. Zaslavsky for fruitful discussions during
the overall process of this work, we also acknowledge useful comments
from the Non linear Dynamics Group at CPT. 
\end{acknowledgement}

\newpage

\newpage
\newpage

\newpage
\newpage

\newpage
\newpage

\newpage

\newpage

\newpage

\newpage
\begin{figure}
\includegraphics[%
  width=8cm]{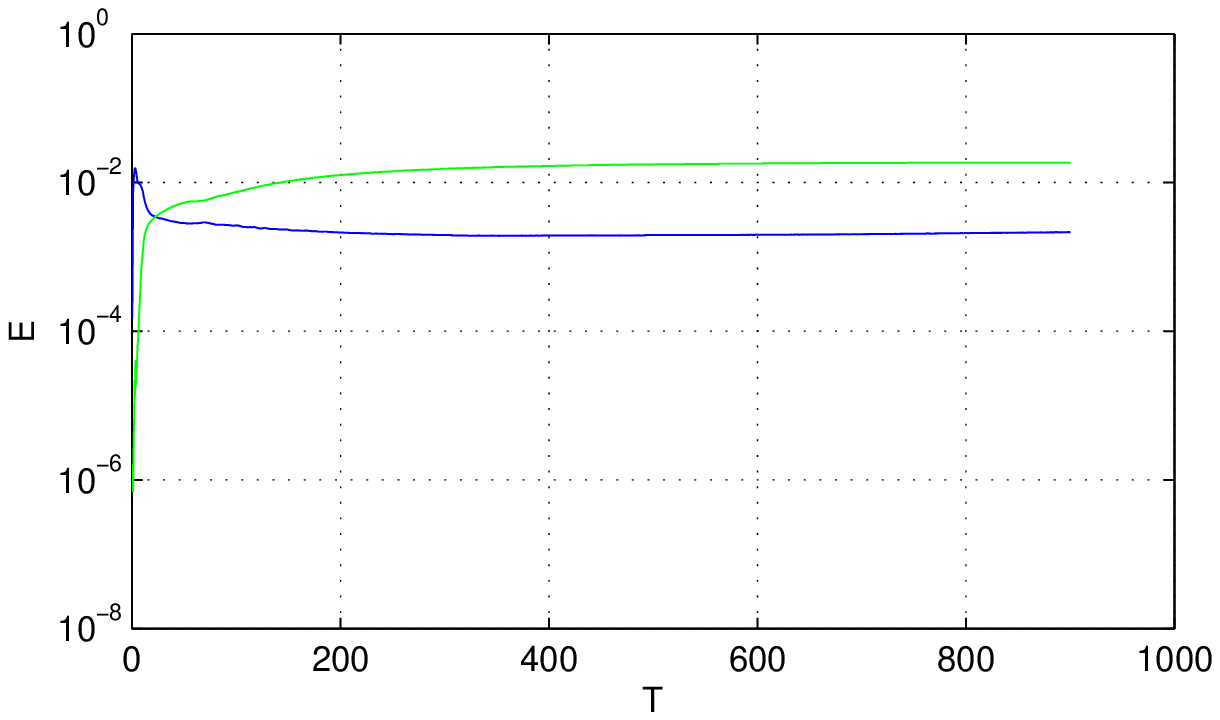}

\caption{Time evolution of kinetic energy (green line) and magnetic energy (blue line). \label{cap:energies_dynamo}}
\end{figure}

\newpage

\end{document}